\documentclass[EPJ,twocolumn]{webofc}
\usepackage[varg]{txfonts}   
\wocname{Atmohead Conference series}
\woctitle{Proceedings of the 2nd Atmohead Conference, May
  19-21, 2014. Padova (Italy) }
\begin{document}
\title{All Sky Camera, LIDAR and Electric Field Meter: auxiliary instruments for the ASTRI SST-2M prototype}
%
%

\author{Giuseppe Leto\inst{1}\fnsep\thanks{\email{gle@oact.inaf.it}} \and
  Ricardo Zanmar Sanchez\inst{1} \and
  Giancarlo Bellassai\inst{1}  \and
  Pietro Bruno\inst{1}  \and
  Maria Concetta Maccarone\inst{2} \and
  Eugenio Martinetti\inst{1} \and
  for the ASTRI Collaboration\inst{3} \and
  the CTA Consortium\inst{4}
}

\institute{INAF - Osservatorio Astrofisico di Catania, Via S. Sofia 78, I-95123 Catania, Italy
  \and
  INAF - IASF Palermo, Via U. La Malfa 153, I-90146 Palermo, Italy
  \and
  http:\/\/www.brera.inaf.it\/astri
  \and
  http:\/\/www.cta-observatory.org
}

\abstract{%
ASTRI SST-2M is the end-to-end prototype telescope of the Italian National Institute of Astrophysics, INAF, designed to investigate the 10-100 TeV band in the framework of the Cherenkov Telescope Array, CTA. The ASTRI SST-2M telescope has been installed in Italy in September 2014, at the INAF observing station located at Serra La Nave on Mount Etna. The telescope is foreseen to be completed and fully operative in spring 2015 including auxiliary instrumentation needed to support both operations and data analysis. In this contribution we present the current status of a sub-set of the auxiliary instruments that are being used at the Serra La Nave site, namely an All Sky Camera, an Electric Field Meter and a Raman Lidar devoted, together with further instrumentation, to the monitoring of the atmospheric and environmental conditions. The data analysis techniques under development for these instruments could be applied at the CTA sites, where similar auxiliary instrumentation will be installed.
}
\maketitle

\section{Introduction}
\label{intro}

The forthcoming international project the Cherenkov Telescope Array, CTA \cite{bib:CTA_2011,bib:CTA_2013} will explore the Very High Energy domain with unprecedented sensitivity and angular resolution. The CTA is designed as an array of many tens of telescopes of three different types with different mirror sizes (Large, Medium, Small) in order to cover the full energy range from few tens of GeV to above 100 TeV. Moreover, two sites are foreseen for CTA, one in the Southern and one in the Northern hemisphere in order to provide full sky coverage.

The INAF contribution to CTA is represented, in its basic form, by the ASTRI program \cite{bib:ASTRI_LaPalombara}, a \emph{Flagship Project} funded by the Italian Ministry of Education, University and Research.
A first goal of the ASTRI project is the realization of an end-to-end prototype of the small-size class of telescopes in a dual-mirror configuration (SST-2M) \cite{bib:ASTRI_ICRC2013} devoted to the investigation of the energy range from a few TeV up to 100 TeV. As a further step, the ASTRI project is a step towards the implementation of an ASTRI/CTA mini-array \cite{bib:ASTRI_CTA_miniarray} composed of seven identical SST-2M telescopes to be placed at the final CTA Southern Site.

In September 2014, the prototype telescope, named ASTRI SST-2M, was installed in Italy at the
'M.G. Fracastoro' observing station of the INAF-Catania Astrophysical Observatory. The station is located in Serra
La Nave (1735 m a.s.l, 37$^\circ$ 41' 05" N Latitude, 14$^\circ$ 58' 04" E Longitude) on Mount Etna \cite{bib:Maccarone_ICRC2013}. The completion of the telescope including camera release is foreseen in Spring 2015.

Being an end-to-end prototype, the ASTRI SST-2M telescope will be fully equipped with all instruments and control systems as if it were at the final CTA site. Together with a dedicated control room and data center \cite{bib:Gianotti_SPIE2014}, all the necessary auxiliary instrumentation will be implemented to support the telescope operations.

Several auxiliary instruments are or are going to be installed at the Serra La Nave (SLN) site, and laboratory/site tests are on going \cite{bib:Leto_AtmoHead2013}.
Part of such instrumentation is devoted to the monitoring of the meteorological and environmental conditions to support the observations, calibration and analysis of the ASTRI SST-2M scientific data \cite{bib:Tosti_SPIE2014}.

In this contribution we refer to three instruments, namely the Color All-Sky Camera, Electric Field Meter and Raman Lidar which are already installed at the SLN site.
We describe the investigated methods to detect clouds from the color all-sky images, the first results from the Electric Field Meter (mainly acting as lightning detector) and the first results in monitoring, with our Lidar system, the Etna volcanic ash that can influence the transparency of the atmosphere.

 \section{The All-Sky Camera}
\label{sec-asc}

The All-Sky Camera (ASC) installed at Serra La Nave site is the SBIG AllSky-340C color fish-eye model \cite{bib:AllSkyCamera}.
The instrument, operating both during daylight and night time, was set to acquire one image (field of view $\sim180 \deg$, 640x480 pixels) every 5 minutes throughout the day. The exposure time is automatically adjusted and ranges from 50 micro seconds to 60 seconds. The analysis of the ASC images will provide a continuous monitoring of the cloud cover for alert purposes.
We are currently investigating two methods to detect clouds: the first method uses a threshold technique plus Principal Component Analysis, the second one is based on star counting via  comparison between expected and observed stars.

\subsection{Thresholding with a Principal Component Analysis}
\label{sec-21}

Thresholding is the simplest method for image segmentation: it looks for a value, a threshold, that can classify the image pixels as either foreground or background; in our case, clouds and clear sky respectively. In this work we are working with colored images which have Red (R), Green (G) and Blue (B) channels.  Different approaches can be found in the literature to application of  thresholding to colored images, for example: finding the threshold on all three channels independently \cite{bib:Kazantzidis_2012}, or transforming the cartesian RGB space into the cylindrical version Hue (H) Saturation (S) and Luminance (HSL) space and applying thresholds on the saturation channel \cite{bib:Souza-Escher2006}. Other approaches (\cite{bib:LiLuYang2011} and references therein) have dropped the green channel and apply thresholds on a R-B or R/B derived channel. None of the above approaches deal with images taken during the night. Since we are mostly interested in the night sky (when the ASTRI SST-2M will observe) we could not a priori argue that the sky is blue and adopt a R-B or R/B channel. We therefore use a Principal Component Analysis (PCA) to discover what is the best mix of channels that better helps us to classify clouds and clear-sky pixels.

\begin{figure}[h!t]
\centering
\includegraphics[width=0.47\textwidth,clip,origin=l]{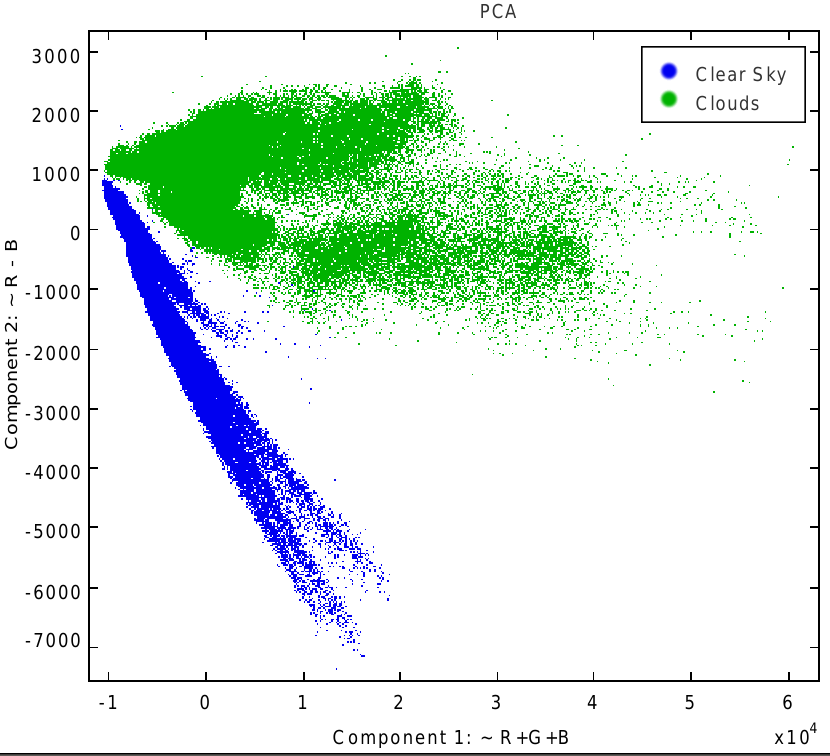}
\caption{The new 2D space with the principal component along the X axis and the second component along the Y axis. The third component turned out to be negligible and is not plotted.}
\label{fig-2DPCA}       
\end{figure}

PCA is a statistical technique that can be used to look for patterns and simplify multidimensional data  \cite{PCABook}. In our case we would like to go from a three dimensional space (RGB) to a single dimension in order to find and apply a single threshold. Operationally speaking, with this method we find the axis along which we have the largest variance (called the principal component), then an orthogonal axis to the first component is found with the second largest variance (called the second principal component), then a third orthogonal axis with the next largest variance, and so on. When applying this method to our images we immediately see that the variance is dominated by the clouds, as shown in Figure \ref{fig-2DPCA}. Note also that clear-sky pixels (marked with blue in Figure \ref{fig-2DPCA}) occupy a distinct region of the space.

\begin{figure}[h!t]
\centering
\includegraphics[width=0.5\textwidth,clip,origin=l]{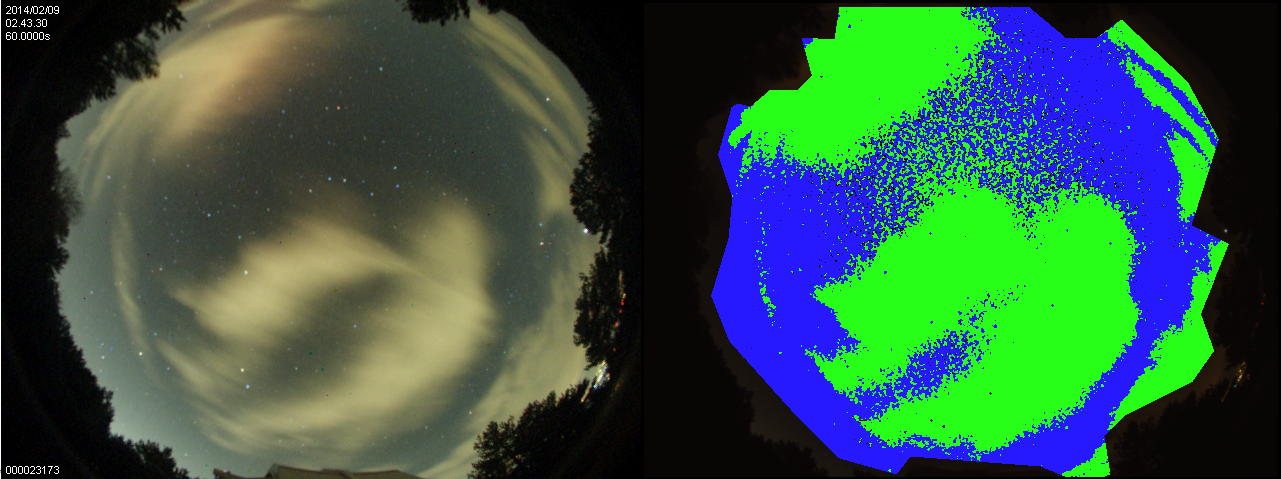}
\caption{\textit{Left}: original image taken during the night with 60 seconds exposure time. \textit{Right}: The result of our thresholding classification, clear-sky pixels are shown in blue and cloudy-sky pixels in green.}
\label{fig-thresholding}       
\end{figure}

Furthermore, the variance along the third component is 3 orders of magnitude smaller than the first component and it can be neglected: the three dimensional RGB space can be reduced to one dimension by rotating the space properly. Based on the observation that different regions are occupied by clouds and clear-sky in the new 2D space, we propose to find the new axis along which we can separate clouds and clear-sky with a single threshold.
The performance of our approach can be evaluated by using the accuracy defined as:
(TP + TN )/(TP + FP + TN + FN), where TP stands for True Positive, TN, True Negative, FP, False Positive, FN, False Negative. We built a database with a large sample of images marking clouds and clear-sky pixels carefully by eye. We use this sample of Ground Truth Images (GTI), to search for the best axis and threshold that allow us to find the best accuracy for each image. An example of the application of the method is presented in Figure \ref{fig-thresholding}. On the left we have the original image, a 60 sec exposure time image taken during the night, on the right we have the result of our classification, with blue marking clear-sky pixels and green, cloudy ones. The classification is obtained by putting the values of the original image in the space of Figure \ref{fig-2DPCA} after having applied the rotation found using GTI in step one, and the threshold to discriminate between clear and cloudy sky.

\subsection{Star counting}
\label{sec-22}

Another approach to detect clouds in images taken during the night is the following: detect stars, compare them with catalogs and infer the cloudiness by comparing the number of stars expected against the number of stars observed.

\begin{figure}[h!t]
\centering
\includegraphics[width=0.5\textwidth,clip,origin=l]{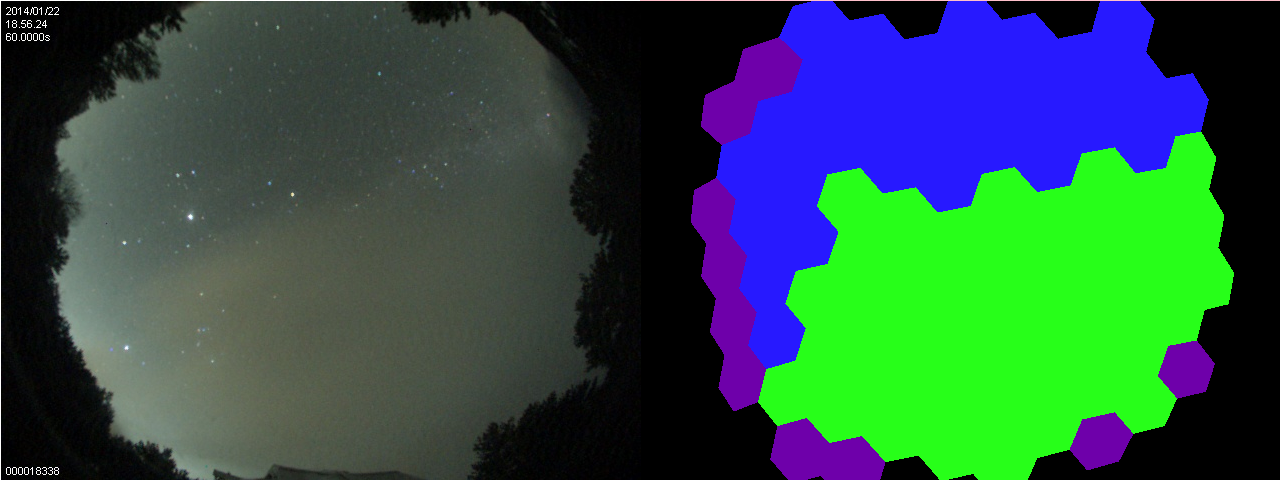}
\caption{Star counting classification: blue is clear, green is cloudy, purple is undetermined (see text)}
\label{fig-starcount}       
\end{figure}

We start with the Hipparcos Catalog \cite{bib:hipparcos} and adapt it to our observatory, e.g. remove stars that can't be observed at our latitude, and that are dimmer than 8th magnitude. We also need to be able to transform pixel coordinates to standard coordinates (RA-DEC). We model our images with a Zenithal Equal Area (ZEA) projection. We allow for distortions using polynomials of up to order 6 and find that our model is good to within $\sim$1 pixel (0.28 deg).
Looking for stars in this kind of images, it is very easy to find a lot of false positives, cosmic rays, moon light reflections,
airplane trails, etc. We therefore iterate the matching process between observed and catalog stars by matching the positions and brightness and removing stars that do not match the brightness within 1 magnitude. The next step is to segment the sky in order to determine if a given tile contains stars or not. For convenience, we have chosen hexagons since they better approximate circles. Figure \ref{fig-starcount} illustrates the result of using this approach.  Clear-sky hexagons, where the ratio between the observed and catalog stars is above a certain threshold, are shown in blue. Cloudy hexagons, which contain a low number of stars, are shown in green, while the purple hexagons on the edges of the field contain insufficient stars for a classification to be made.

\section{The Raman LIDAR - AMPLE}
\label{sec-lidar}
The LIDAR (LIght Detection And Ranging) called AMPLE has been in operation since October 2013 at the SLN site within a project in collaboration with INGV-Catania (National Institute of Geophysics and Vulcanology) \cite{bib:scolloample}.

\begin{figure} [h!]
\vspace*{0cm}
\centering
\includegraphics[width=0.47\textwidth,clip,origin=l]{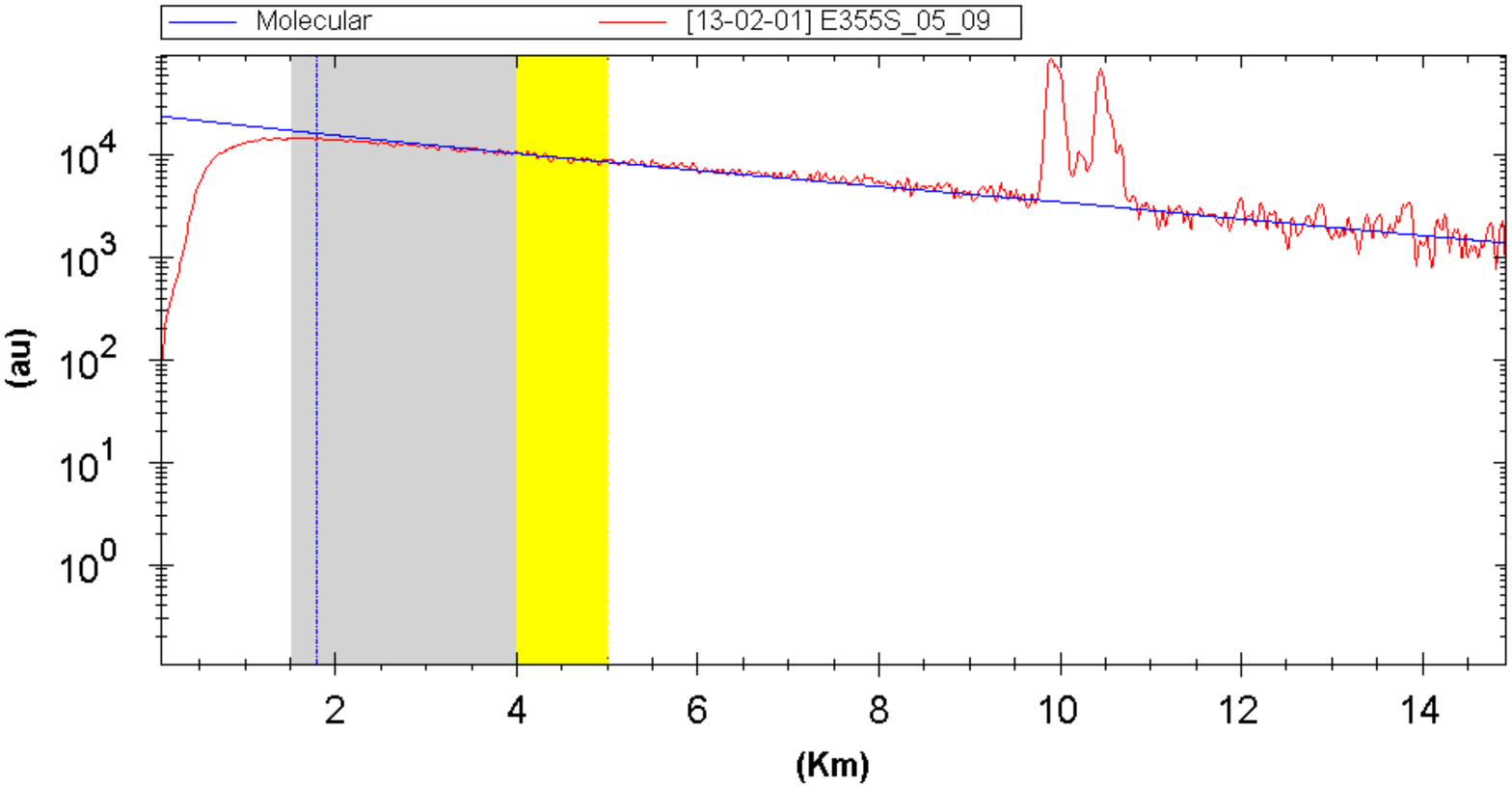}
\vspace*{-1.0cm}
\caption{AMPLE LIDAR sample measure, intensity of backscattered signal (arbitrary units) as a function of the distance.}
\label{lidarmeas}       
\end{figure}

The LIDAR AMPLE is equipped with elastic channels at 355 nm with a polarization option available and a Raman channel at 386nm. The system can measure backscattered light from a few hundred meters up to 15 km (see figure~\ref{lidarmeas}). AMPLE  is
controlled by a dedicated software and can perform both pointed and scanned observations by means of a programmable interface. AMPLE can also be controlled by a remote observer.
Lidar measurements will provide data about the presence of dust and aerosols in the atmosphere during ASTRI SST-2M observation. 

\section{The Electric Field Meter}
\label{sec-efm}

\begin{figure}[h!t]
\centering
\includegraphics[width=0.26\textwidth,clip,angle=-90,origin=l]{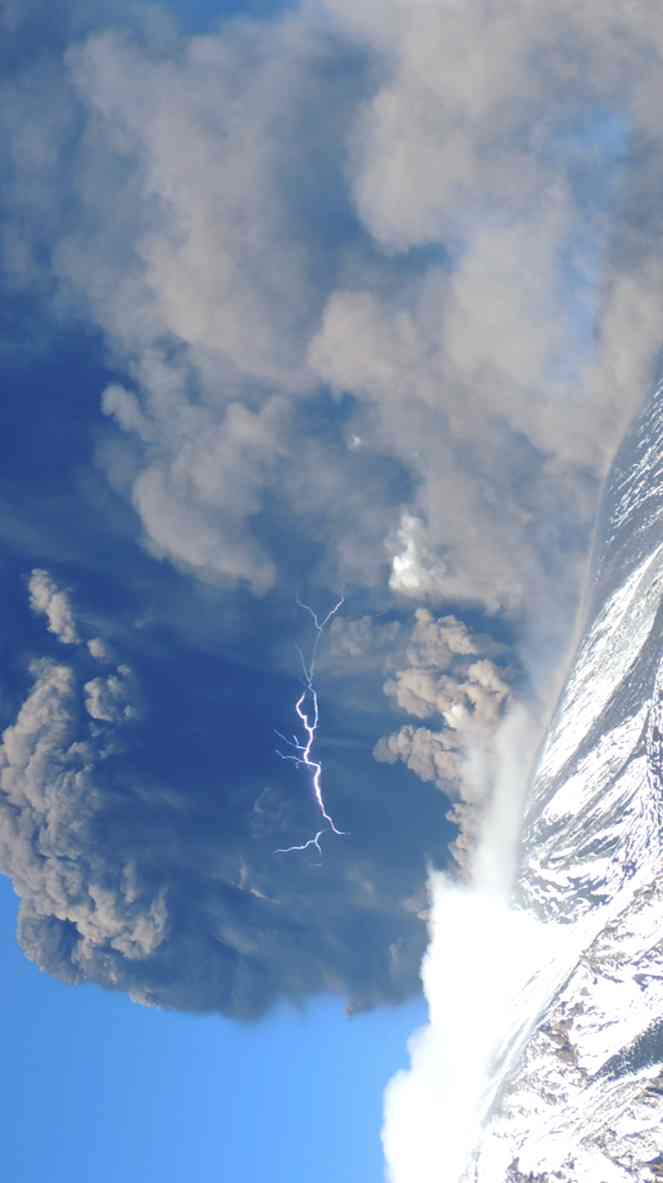}
\caption{Lightning event during the lava fountain episode on Nov 23, 2013.}
\label{lightning}       
\end{figure}

In order to monitor the lightning activity, an Electric Field Meter (EFM) \cite{bib:EFM}, developed by a research group of the INAF Radio Astronomy Institute (IRA) in Bologna, was deployed at the SLN site during Fall 2013 \cite{bib:EFM,bib:Leto_AtmoHead2013}. The EFM will be used for monitoring atmospheric electric field variations and to issue alerts to the telescope control system. If used in an array configuration, EFM is also able to give detailed information about the direction of the lightning activity and the approaching speed.
During the initial EFM testing period an Etna Volcano eruption event gave us the opportunity to characterize the instrument. In fact, during the few hours of an active lava fountain a number of lightning events occurred.  The sky was clear and no other obvious source of lightning activity was present.  
Events were recorded in a video (see Figure~\ref{lightning} shows a photograph as an example) and the exact timing could be determined. The video also enabled us to locate spatially the lava fountain with precision because it was a vertical column on top of the New South-East crater. The distance of the crater from the observing site is 6.4 km. Since the video shows that the lightning was concentrated within the column, we could estimate that the EFM is able to detect lightning events at least at a distance of 6.4$\pm$0.3 km. \\

\begin{figure}[h!t]

\vspace*{-1.0cm}
\centering
\includegraphics[width=0.33\textwidth,clip,origin=l,angle=-90]{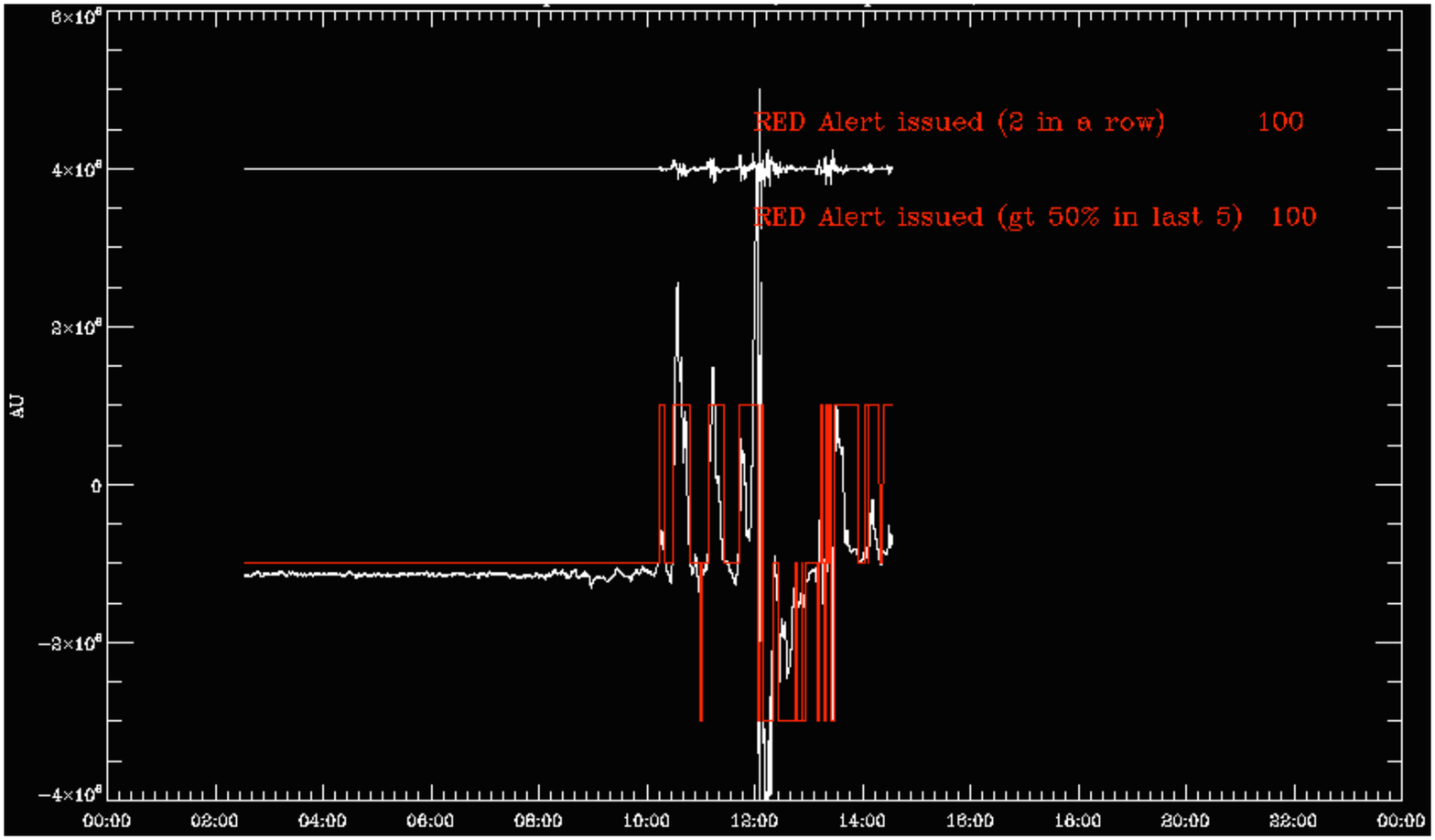}
\vspace*{-0.5cm}
\caption{Alarm triggered using EFM signal on Jan 5, 2014. Upper white line represents the derivative, lower white line the signal, red line indicate where alert was issued.}
\label{alarm}       
\end{figure}

Data taken with the EFM during the November 2013 event have been used to develop a strategy to identify lightning events. The unperturbed EFM signal level and its typical uncertainty varies with time as a function of ambient and weather conditions. This variation is slow and can be monitored with the first derivative of the signal.
 When a sudden variation due to lightning appears, the procedure detects it as a strong deviation from the actual unperturbed values and sets a variable to "true"; it is possible to choose whether to send an immediate alarm or wait for a second and/or a third "true" to issue the alarm to the atmospheric monitoring software (Figure~\ref{alarm}) in order to stop the telescope observations. \\

\section{Conclusions}
\label{concl}

The ASTRI SST-2M telescope, an end-to-end prototype of the Small Size class Telescope for the CTA, has been installed in Italy at the Serra La Nave site. The completion of the telescope is foreseen in Spring 2015 together with all the instrumentation necessary to support the telescope operations. Among the planned auxiliary instruments, the All-Sky Camera, an EFM and a LIDAR are already active on site and new methods are currently under development to extract, at a proper high level of significance, the information needed to monitor part of the meteorological, atmosphere and weather conditions. The early results obtained with these methods are very encouraging and work is in progress to optimize them and improve statistics. \\

\vspace*{0.5cm}
\footnotesize{{\bf Acknowledgment:}{This work was partially supported by the ASTRI "Flagship Project" financed by the Italian Ministry of Education, University, and Research (MIUR) and led by INAF, the Italian National Institute of Astrophysics. We also acknowledge partial support by the MIUR 'Bando PRIN 2009' and TeChe.it 2014 Special Grants. We gratefully acknowledge support from the agencies and organizations listed under Funding Agencies at the CTA web site http:\/\/www.cta-observatory.org. INAF-OACT also acknowledges the support for atmospheric monitoring by PO Italia Malta 2007-2013 with VAMOS SEGURO project, A1.2.3-62 in the framework of the collaboration of OA-Catania with the INGV-Catania .}} \\

%
%
%

\end{document}